\documentclass[11pt]{article}
\usepackage{latexsym}  
\usepackage{amssymb}
\usepackage{graphicx}
\usepackage{amsmath}

\begin{document}
\hspace*{11cm}  {OU-HET-662/2010}

\begin{center}
{\Large\bf Charged Lepton Mass Spectrum }\\
{\Large\bf and a Scalar Potential Model}

\vspace{5mm}
{\bf Yoshio Koide}

{\it Department of Physics, Osaka University,  
Toyonaka, Osaka 560-0043, Japan} \\
{\it E-mail address: koide@het.phys.sci.osaka-u.ac.jp}
\end{center}

\begin{abstract}
Stimulated by a recent work by Sumino, on the basis of 
a model in which effective Yukawa coupling constants
are described by vacuum expectation values of a scalar 
$\Phi$ with $3\times 3$ components, a possible form of 
the scalar potential 
$V(\Phi)$ which can well describe the observed charged 
lepton mass spectrum is investigated without 
referring to a specific flavor symmetry model. 
A general relation among eigenvalues of 
$\langle \Phi\rangle$ is derived without referring to
an explicit form of $V(\Phi)$. 
\end{abstract}

\vspace{3mm}

{\large\bf 1 \ Introduction}

In the standard model, mass spectra of quarks and leptons 
are originated in the Yukawa coupling constants which are
fundamental constants in the theory.
On the other hand, so for, many phenomenological relations 
in the observed mass spectra and mixings have been known.
Therefore, it is unlikely that there are so many fundamental 
constants in the nature.
Then, one may consider that the Yukawa coupling 
constants are apparent constants in an effective theory. 
Previously, Froggatt and Neilsen \cite{Froggatt} have 
proposed a model in which masses and mixings are 
explained by a multiplicative structure 
$(\langle \phi \rangle/\Lambda)^n$ of a vacuum expectation 
value (VEV) $\langle \phi \rangle$ of a scalar $\phi$.
However, in this paper, we will try to understand the hierarchical
structures of the flavors not by U(1) flavor charges 
like a Froggatt-Neilsen model, but by a VEV structure 
$\langle \Phi \rangle$ of a scalar $\Phi$ with
$3\times 3$ components. 
For example, in order to explain a charged lepton
mass relation \cite{Koidemass}
$$
K \equiv 
\frac{m_e +m_\mu +m_\tau}{(\sqrt{m_e} +\sqrt{m_\mu}
+\sqrt{m_\tau})^2} = \frac{2}{3} ,
\eqno(1.1)
$$
the author \cite{K-mass90} has previously proposed a simple 
form of the scalar potential $V(\Phi)$ under an assumption that 
the charged lepton masses $m_{ei}$ are given by 
$m_{ei} \propto v_i^2$, where $(v_1, v_2, v_3)$ are eigenvalues 
of $\langle\Phi\rangle$.
The relation (1.1) is well satisfied with the order of $10^{-5}$ 
for the observed pole masses \cite{PDG08}, i.e. 
$K^{pole}=(2/3)\times (0.999989 \pm 0.000014)$. 
However, in conventional mass matrix models, ``masses" 
mean not ``pole masses", but ``running masses".
The formula (1.1) is, at most, valid with the order of 
$10^{-3}$ for the running masses.

Recently, a possible solution of this problem has been 
proposed by Sumino \cite{Sumino09PLB,Sumino09JHEP}:
He considers that a flavor symmetry is gauged, and 
radiative corrections due to photon to the charged 
lepton masses are exactly canceled by those due to  
flavor gauge bosons.
Therefore, as far as the charged lepton masses are
concerned, the running mass values are exactly equal to 
the pole mass values. 
(Hereafter, we will refer to this mechanism as Sumino's mechanism.)
Moreover, as a byproduct of his model, he has obtained
a charged lepton mass relation \cite{Sumino09JHEP}
$$
m_\mu^{3/2} +\sqrt{m_e m_\mu m_\tau} = \sqrt{ m_e m_\tau}
(\sqrt{m_e}+\sqrt{m_\tau}) .
\eqno(1.2)
$$
(We also refer to the relation (1.2) as Sumino's 
relation on the charged lepton masses.)
Then, we can completely determine the charged lepton 
mass spectrum by using two relations (1.1) and (1.2)
simultaneously.
The predicted value of 
the electron mass $m_e$ is in fairly good agreement 
with the observed one. 

In the present paper, stimulated by the Sumino relation
(1.2), we investigate a possible scalar potential form 
of $\Phi$.
However, we do not refer to the Sumino mechanism, 
so that we do not discuss whether the flavor symmetry is 
gauged or global. 
We discuss the charged lepton mass relations not with the 
order of $10^{-5}$ but with the order of $10^{-3}$. 
Since our interest is only in the VEV values 
$(v_1, v_2, v_3)$ of the eigenvalues of 
$\langle\Phi\rangle$, we denote the scalar potential 
$V(\Phi)$ in terms of $v_i$ as follows:
$$
V_S=\lambda_1 (v_1^2 v_3^2 + v_2^2 v_2^2 + v_3^2 v_1^2)
+\lambda_2 (v_1 v_3 + v_2 v_2 + v_3 v_1)^2
$$
$$
+\lambda_3 (v_1^4+v_2^4+v_3^4) +\lambda_4 (v_1^2+v_2^2+v_3^2)^2,
\eqno(1.3)
$$
where we have denoted only typical terms which are presumed 
from the original Sumino potential \cite{Sumino09JHEP}. 
In the original Sumino model \cite{Sumino09JHEP}, the potential 
$V(\Phi)$ has been given only by $\lambda_1$ and $\lambda_3$ 
terms.
In contrast to the Sumino model, in order to derive the 
Sumino relation (1.2), the author  \cite{Koide10PLB} has 
recently proposed a model based on a so-called 
``supersymmetric yukawaon" model, where a superpotential 
$W(\Phi)$ is practically given by the $\lambda_1$, $\lambda_2$ 
and $\lambda_4$ terms, and the $\lambda_3$ term is forbidden 
by $R$-charge conservation.
In any cases, the Sumino relation (1.1) is derived from 
the first term in the potential (1.3), so that the first term 
must be a dominant term in the model, and other terms play 
a role in giving small corrections to the Sumino relation (1.2).
Hereafter, we refer to the potential (1.3) as an extended 
Sumino potential.

The Sumino model is based on a flavor symmetry U(9) 
which is reduced to U(3)$\times$O(3). 
On the other hand, the potential form (1.3) in the supersymmetric 
yukawaon model \cite{Koide10PLB} is based on a U(3) flavor symmetry 
(however, by assuming many subsidiary fields). 
In this paper, we will also not refer to an explicit model  
for the potential form (1.3) from a practical point of view. 
Since the purpose of the present paper is to discuss only a 
substantial framework of the extended Sumino potential (1.3), 
we deal with $\Phi$ as if it is a plain $3\times 3$ matrix and 
we do not discuss what flavor symmetry is supposed in a model
concerned and to what multiplet the field $\Phi$ belongs. 
For example, when we consider a diagonal form of 
$\langle\Phi\rangle$ and a form 
$\langle\bar{\Phi}\rangle \equiv T \langle\Phi\rangle T$
as follows 
$$
\Phi = \left(
\begin{array}{ccc}
v_1 & 0 & 0 \\
0 & v_2 & 0 \\
0 & 0 & v_3
\end{array} \right) , \ \ \ 
\bar{\Phi} = \left(
\begin{array}{ccc}
v_3 & 0 & 0 \\
0 & v_2 & 0 \\
0 & 0 & v_1
\end{array} \right) , 
\eqno(1.4)
$$
by assuming a transformation matrix 
$$
T= \left(
\begin{array}{ccc}
0 & 0 & 1 \\
0 & 1 & 0 \\
1 & 0 & 0
\end{array} \right) ,
\eqno(1.5)
$$
(here and hereafter, for simplicity, we denote the VEV matrix 
$\langle\Phi\rangle$ as $\Phi$ simply), the $\lambda_1$, $\cdots$
and $\lambda_4$ terms in the potential (1.3) can be expressed as
${\rm Tr}[\Phi \bar{\Phi} \Phi \bar{\Phi}]$, 
${\rm Tr}[\Phi \bar{\Phi}] {\rm Tr}[\Phi \bar{\Phi}]$, 
${\rm Tr}[\Phi {\Phi} {\Phi} {\Phi}]$, and
${\rm Tr}[\Phi {\Phi}] {\rm Tr}[\bar{\Phi} \bar{\Phi}]$, 
respectively. 
On the other hand, if we choose a flavor basis in which 
a form of $\langle\Phi\rangle$ is given by
$$
\Phi = \left(
\begin{array}{ccc}
0 & 0 & v_1 \\
0 & v_2 & 0 \\
v_3 & 0 & 0
\end{array} \right) , \ \ \ 
{\Phi}^T = \left(
\begin{array}{ccc}
0 & 0 & v_3 \\
0 & v_2 & 0 \\
v_1 & 0 & 0
\end{array} \right) , 
\eqno(1.6)
$$
we can express those terms in (1.3) as 
${\rm Tr}[\Phi {\Phi} \Phi {\Phi}]$, 
${\rm Tr}[\Phi {\Phi}] {\rm Tr}[\Phi {\Phi}]$, 
${\rm Tr}[\Phi {\Phi}^T {\Phi} {\Phi}^T]$, and
${\rm Tr}[\Phi {\Phi}^T] {\rm Tr}[{\Phi} {\Phi}^T]$, 
respectively. 
In any cases, in order to build a realistic model, we 
will be obliged to give an explicit model of the flavor 
symmetry and to introduce some additional fields. 
However, the purpose of the present paper is to 
investigate a hint for a further extension of the 
Sumino model, so that we will not refer to  
such a realistic model. 

In the Sumino model \cite{Sumino09JHEP}, 
the relation (1.2) has been derived on the condition
that the relation (1.1) is exactly satisfied. 
On the other hand, in the supersymmetric yukawaon 
model \cite{Koide10PLB}, the relation (1.1) is satisfied 
only as an approximate relation for the running masses. 
Another interest in the present paper is to see what happens
in the charged lepton mass spectrum when we put
a potential term $V_K$ which leads to the relation (1.1)
in addition to the Sumino potential (1.3).
In the present paper, we adopt the following form 
\cite{K-mass90} as $V_K$:
$$
V_K = \mu^2 {\rm Tr}[\Phi\Phi] + \lambda_K {\rm Tr}[\hat{\Phi} \hat{\Phi}]
({\rm Tr}[\Phi])^2 
$$
$$
=\mu^2 (v_1^2+v_2^2+v_3^2) + \lambda_K \left[v_1^2+v_2^2+v_3^2 
-\frac{1}{3}(v_1+v_2+v_3)^2\right](v_1+v_2+v_3)^2 ,
\eqno(1.7)
$$
where $\hat{\Phi}$ is a traceless matrix which is defined
as $\hat{\Phi}= \Phi -\frac{1}{3} {\rm Tr}[\Phi]$ in the
expression (1.4). 
[In Ref.\cite{K-mass90}, the form (1.7) has been given by assuming
that the mass term (the $\mu^2$ term) is given in terms of 
a nonet (${\bf 8}+{\bf 1}$) form of U(3), while the dimension 
4 term (the $\lambda_K$ term) is given only in terms of a form 
``octet-octet".]
In the next section, we will investigate a possible VEV 
spectrum of $\Phi$ under the potential $V=V_K+V_S$. 
Since we have 4 parameters in the potential (1.3), we
cannot predict the charged lepton mass spectrum under 
the general form (1.3) of $V_K$. 
However, when we build an explicit model, we can drop some 
of the terms in the potential (1.3) by assuming some  
additional symmetries. 
The purpose is to see whether we can obtain a more
plausible result by dropping some of the terms in the 
potential (1.3) or not. 
Then, the result will provide us a promising clue to 
a realistic lepton mass matrix model.

\vspace{3mm}

{\large\bf 2 \ Minimizing conditions of the potential}

In this section, we investigate minimizing conditions of 
$V=V_K+V_S$.
From the explicit form of $V_K$, (1.7), but without assuming  
an explicit form of $V_S$, we obtain the following conditions
$$
\frac{\partial V}{\partial v_i} = 2 v_i \left[ A + 
\frac{1}{v_i} (B + B_i) \right] = 0 ,
\eqno(2.1)
$$
where $i=1,2,3$ and
$$
A=\mu^2 + \lambda_K (v_1+v_2+v_3)^2 ,
\eqno(2.2)
$$
$$ 
B = \lambda_K (v_1+v_2+v_3) \left[ v_1^2+v_2^2+v_3^2
-\frac{2}{3} (v_1+v_2+v_3)^2 \right] ,
\eqno(2.3)
$$
$$
B_i = \frac{1}{2} \frac{\partial V_S}{\partial v_i} .
\eqno(2.4)
$$
Hereafter, we investigate only a case with $v_i \neq 0$ 
for all $i=1,2,3$.

If the extended Sumino potential (1.3) is absent, 
i.e. $\lambda_1=\lambda_2=\lambda_3=\lambda_4=0$, 
the condition $A+B/v_1=A+B/v_2=A+B/v_3=0$ leads to
$A=B=0$, so that we obtain
$$
K\equiv \frac{v_1^2+v_2^2+v_3^2}{(v_1+v_2+v_3)^2} =
\frac{2}{3} ,
\eqno(2.5)
$$
together with
$$
(v_1+v_2+v_3)^2 = - \frac{\mu^2}{\lambda_K} .
\eqno(2.6)
$$
The result (2.5) means the charged lepton mass relation
(1.1).
However, when the Sumino potential (1.3) exists, the 
relation (2.5) is not satisfied as we see below.

When $V_S$ exists, independently of the explicit form of
$V_S$, we obtain three conditions 
$$
-A = \frac{1}{v_1}(B+B_1) =\frac{1}{v_2}(B+B_2) =
\frac{1}{v_3}(B+B_3) .
\eqno(2.7)
$$
From Eq.(2.7), we obtain
$(B+B_1)/v_1=(B+B_3)/v_3$, i.e.
$$
B =\frac{1}{v_1-v_3}(v_3B_1-v_1 B_3).
\eqno(2.8)
$$
By using a general formula $k=(c_1 a+ c_2 b)/(c_1+c_2)$ for 
a given relation $k=a=b$, we can write
$A$ as follows:
$$
-A= \frac{1}{v_1+v_3} \left( v_1 \frac{B+B_1}{v_1}
+ v_3 \frac{B+B_3}{v_3} \right) = \frac{1}{v_1+v_3} 
(2 B+B_1+B_3) .
\eqno(2.9)
$$
When we eliminate $B$ from $(B+B_2)/v_2=(2 B+B_1+B_3)/(v_1+v_3)$ 
by using Eq.(2.8), we can obtain a general formula for $v_i$
$$
(v_3-v_2) B_1 +(v_2-v_1)B_3 +(v_1-v_3) B_2 =0 .
\eqno(2.10)
$$
Also, from Eq.(2.9), we obtain
$$
A= - \frac{B_1-B_3}{v_1-v_3} .
\eqno(2.11)
$$
Hereafter, we will use the conditions (2.8), (2.10) and (2.11)
instead of the three conditions $\partial V/\partial v_i =0$ 
as minimizing conditions of $V=V_K+V_S$.
We will refer to the conditions (2.11), (2.8) and (2.10) 
as the conditions A, B and C, respectively.
Note that the condition C is satisfied independently of the 
values of $A$ and $B$ which depend on the parameters
$\mu^2$ and $\lambda_K$ in $V_K$.

For the explicit form of $V_S$, (1.3), we obtain
$$
\lambda_1 S(v_1, v_2, v_3) + \lambda_2 (v_2^2 + 2 v_1 v_3)
(2 v_2 -v_1-v_3) 
$$
$$
+\lambda_3 [ (2 v_2^2 -v_1^2-v_3^2)v_2  -S(v_1,v_2,v_3)] = 0 ,
\eqno(2.12)
$$
where the factor $S(v_1,v_2,v_3)$ is defined by
$$
S(v_1,v_2,v_3) =v_2^3 + v_1 v_3 v_2 -v_1 v_3 (v_1+v_3) ,
\eqno(2.13)
$$
and $S(v_1,v_2,v_3)=0$ gives the Sumino mass relation (1.2).

Note that the $\lambda_4$ term (and also a term with a form
$(v_1+v_2+v_3)^n$) does not affect the general formula (2.10).
However, this does not mean that the $\lambda_4$ term does 
not affect the minimizing conditions.
The $\lambda_4$ term contributes to the factors $B_i$
as follows:
$$
\begin{array}{l}
B_1=2 \lambda_1 v_1 v_3^2 + 2 \lambda_2 (v_2^2+2 v_1 v_3) v_3 
+2 \lambda_3 v_1^3 + 2 \lambda_4 (v_1^2+v_2^2+v_3^2) v_1, \\
B_3=2 \lambda_1 v_1^2 v_3 + 2 \lambda_2 (v_2^2+2 v_1 v_3) v_1 
+2 \lambda_3 v_3^3 + 2 \lambda_4 (v_1^2+v_2^2+v_3^2) v_3, \\
B_2=2 \lambda_1 v_2^3 + 2 \lambda_2 (v_2^2+2 v_1 v_3) v_2 
+2 \lambda_3 v_2^3 + 2 \lambda_4 (v_1^2+v_2^2+v_3^2) v_2 . 
\end{array}
\eqno(2.14)
$$
Therefore, the $\lambda_4$ term contributes to the condition 
B, Eq.(2.11), as follows:
$$
A=2 [\lambda_1 v_1 v_3 + \lambda_2 (v_2^2+ 2 v_1 v_3) 
-\lambda_3 (v_1^2+v_1 v_3 +v_3^2) -\lambda_4 (v_1^2+v_2^2 +v_3^2) ] .
\eqno(2.15)
$$
In contrast to the condition A, (2.15), the $\lambda_4$ term
does not contribute to the condition B, (2.8), i.e.
$$
B = - 2(v_1+v_3) [\lambda_1 v_1 v_3 +
\lambda_2 (v_2^2 +2 v_1 v_3) -\lambda_3 v_1 v_3 ] .
\eqno(2.16)
$$
  
Since we have no information on the values $A$ and $B$ which
are defined by Eqs.(2.2) and (2.3), respectively, what we can
use to investigate the structure of $V_K$ is only the 
condition C, (2.12).
For reference, we denote numerical values of Eq.(2.12):
$$
0.00127398\, \lambda_1 -0.0453018\, \lambda_2 
-0.198278\, \lambda_3=0.
\eqno(2.17)
$$
Here, we have used the observed pole mass values \cite{PDG08} 
and the numerical values have been given in the unit 
of $v_0^3$ [$v_0^2\equiv v_1^2+v_2^2+v_3^2$, and
$(v_1,v_2, v_3)/v_0 =(0.0164734,\, 0.236879, \, 
0.971400)$]. 
If we assume an exact Sumino mechanism (cancellation 
between photon and family gauge boson diagrams), the 
relation (2.17) must be satisfied exactly.
Even if we does not require such a cancellation, the 
numerical relation (2.17) must approximately be satisfied 
because we know that the deviation from the mass formula 
(1.1) is only $10^{-3}$ for running masses at a typical 
energy scale ($\mu \sim 10^3 - 10^{16}$ GeV) .

We readily find that there is no solution of 
$(\lambda_1, \lambda_2, \lambda_3)$ except for 
a hierarchical case.
For example, if we suppose a case $\lambda_2=0$,
which is a case in the Sumino model \cite{Sumino09JHEP}, 
we obtain $\lambda_3/\lambda_1=0.006422$, and if we suppose
a case $\lambda_3=0$, which is a case in the yukawaon model
\cite{Koide10PLB}, we obtain $\lambda_2/\lambda_1=0.028122$.
Thus, we cannot have a solution with the same order of 
$\lambda_i$.

In this paper, we will investigate a case with a more 
plausible value of $\lambda_i/\lambda_j$.
For such the purpose, we will discuss the remaining conditions 
A and B in the next section.

\vspace{3mm}

{\large\bf 3 \ Possible form of $V_S$} 

In this section, we discuss the conditions A, (2.15), 
and B, (2.16), in addition to the condition C, (2.12).
First, we investigate a possibility of $B=0$, so that 
the condition B is given by
$$
\lambda_1 v_1 v_3 +
\lambda_2 (v_2^2 +2 v_1 v_3) -\lambda_3 v_1 v_3 =0.
\eqno(3.1)
$$
In the present paper, the relation (2.5) is not always
exactly satisfied. 
For convenience, let us define the following parameter
$\xi$:
$$
K\equiv \frac{v_1^2+v_2^2+v_3^2}{(v_1+v_2+v_3)^2} =
\frac{2}{3} (1+\xi) ,
\eqno(3.2)
$$
so that the parameter $B$ is expressed as
$$
B =\frac{2}{3} \lambda_K (v_1+v_2+v_3)^3 \xi .
\eqno(3.3)
$$
For example, the value of $\xi$ is  $\xi=(0.011 \pm 0.014)
\times 10^{-3}$ for pole masses, while $\xi=(1.95\pm 0.02)
\times 10^{-3}$ and $\xi=(2.30\pm 0.02) \times 10^{-3}$ at
$\mu=10^{3}$ GeV and $\mu=10^{12}$ GeV, respectively,
for running masses in a SUSY model with $\tan\beta=10$. 
(See Ref.\cite{Koide10PLB}.) 
If we accept the Sumino mechanism, the 
value of $\xi$ must exactly be zero, so that 
we can regard the value of $B$ as $B=0$.  
(Of course, the case $\xi=0$ cannot be derived from
the present formulation.)
Then, we can use the second condition (3.1) in addition
to the first condition (2.12).
We have two conditions for three parameters $\lambda_1$, 
$\lambda_2$ and $\lambda_3$, so that we can completely 
determine the relative rations among $\lambda_i$.  
When we use the pole mass values for $m_{ei}$, we  
obtain numerical results
$$
\frac{\lambda_2}{\lambda_1}= -0.173249, \ \ \ 
\frac{\lambda_3}{\lambda_1}= 0.046009.
\eqno(3.4)
$$
Again, we obtain a hierarchical result for $\lambda_i$.
However, it is not likely that these values (3.4) can be
understood by considering a specific model.

Next, we investigate a case with $A=0$. 
By way of trial, we assume that the $\mu^2$ term 
in $V_K$ is always canceled by the term 
$\lambda_K ( v_1+v_2+v_3)^2$, i.e. Eq.(2.6).
Then, the condition A, (2.15), with the value $A=0$ 
leads to a condition
$$
\lambda_1 v_1 v_3 + \lambda_2 (v_2^2+ 2 v_1 v_3) 
-\lambda_3 (v_1^2+v_1 v_3 +v_3^2) -\lambda_4 
(v_1^2+v_2^2 +v_3^2) =0 .
\eqno(3.5)
$$
The condition (3.5) has a special solution
$\lambda_2/\lambda_1=\lambda_3/\lambda_1= - 
\lambda_4/\lambda_1=-1/3$ independently of 
values of $v_i$.
However, the solution does not satisfy 
the condition (2.12), so that the solution is ruled out.
Since the condition (3.5) contains four parameters 
$\lambda_1$, $\cdots$, $\lambda_4$, we cannot determine 
the relative ratios among $\lambda_i$ 
unless we put an ansatz in addition to the conditions
(2.12) and (3.5). 

Since the $\lambda_1$ term is a main term in the Sumino 
relation, we consider $\lambda_1 \neq 0$.  
Then, we eliminate $\lambda_1$ from (2.12) and (3.5), 
we obtain
$$
\lambda_2 (v_2^2+2 v_1 v_3)(v_1 v_3-v_2^2) v_2 
+\lambda_3 [(v_1^2+v_3^2) S 
+ v_1 v_3 v_2 (2 v_2^2-v_1^2-v_3^2)]
$$
$$ 
+\lambda_4 (v_1^2+v_2^2+v_3^2) S =0 ,
\eqno(3.6)
$$
where $S=S(v_1,v_2,v_3)$ defined by Eq.(2.13).
A numerical relation of (3.6) for the pole mass values
is expressed as $-0.83719\lambda_2 -1.95002\lambda_3
+1.27398\lambda_4=0$, i.e.
$$
\lambda_4 =  \frac{1}{1.521735}\lambda_2 + 
1.530652\, \lambda_3 .
\eqno(3.7)
$$
In the yukawaon model \cite{Koide10PLB}, the $\lambda_3$
term is forbidden because of $R$ charge conservation, so
that the condition (3.7) suggests $\lambda_2 = (2/3)\lambda_4$,
which is a likely case. 
[In the yukawaon model, the $\lambda_1$ term includes 
a contribution from another mechanism, so that the value of
$|\lambda_1|$ can be much larger than the values 
$|\lambda_i|$ ($i=2,3$).
For numerical fitting on the case with $\lambda_2 = 
(2/3)\lambda_4$, see Ref.\cite{Koide10PLB}, where 
the numerical fitting for running masses at several energy 
scales has been demonstrated.]
Also, we may speculate a solution $\lambda_3=(2/3)
\lambda_4$ for a case $\lambda_2=0$.
Anyhow, the relation (3.7) suggests an interesting
constraint  on the coefficients $\lambda_i$:
$$
\lambda_3 -\frac{2}{3}\lambda_4 + \left(\frac{2}{3}\right)^2
\lambda_2 = 0 .
\eqno(3.8)
$$
[Note that the numerical relation (3.7) does not need to 
be exactly satisfied because we deal with running masses
$m_{ei}(\mu)$ whose values are dependent on model 
parameters such as $\mu=\Lambda$, $\tan\beta$, and so on.]

For a case with $A=B=0$, we cannot find
a suggestive relation among $\lambda_i$ because the
conditions are too tight.

\vspace{3mm}

{\large\bf 4 \ Concluding remarks} 

In conclusion, by stimulating Sumino's charged lepton 
mass relation, (1.2), we have investigated a possible 
scalar potential form of the field $\Phi$, which plays 
a role in giving the charged lepton mass spectrum via  
$m_{ei} \propto v_i^2$ ($v_i$ are eigenvalues
of $\langle\Phi\rangle$).
We have obtained a general formula for $v_i$, Eq.(2.10),  
independently of a form of the potential $V(\Phi)$.
For an explicit form of $V_S$, (1.3), we obtain the relation
(2.12).
For a model with $\lambda_2=\lambda_3=\lambda_4=0$, we obtain
$S(v_1,v_2,v_3)=0$, so that we get the Sumino relation (1.2)
which is fairly satisfied with realistic charged lepton masses
(for both cases of pole masses and running masses).
For the pole masses, the condition (2.12) is numerically expressed
as Eq.(2.17). 
As seen in Eq.(2.17), we must consider that the $\lambda_2$ and
$\lambda_3$ terms must be considerably suppressed compared with
the first term ($\lambda_1$ term) in order to protect the Sumino 
relation $S \simeq 0$ from destroying badly.

Also, we have obtained relations (2.11) (Condition A) and (2.8) 
(Condition B) which are dependent on $V_K$, and which are given
by Eqs.(2.15) and (2.16) for the explicit form of $V_K$, (1.3). 
The condition (2.16) is rewritten as
$$
\lambda_1 -\lambda_3 +\lambda_2 \frac{v_2^2+2 v_1 v_3}{v_1 v_3}
=-\frac{1}{3}\lambda_K \xi \frac{(v_1+v_2+v_3)^3}{
(v_1+v_3) v_1 v_3} ,
\eqno(4.1)
$$
which is numerically expressed as
$$
\lambda_1 -\lambda_3 + 5.5065\, \lambda_2  = 
38.738\, \lambda_K \xi , 
\eqno(4.2)
$$
where, for convenience, we have used the pole mass values 
for $v_i$ although the value of $\xi$ for pole mass values
is zero.
Since the observed value of $\xi$ is $\xi \sim 10^{-3}$
for the running masses, the left hand side of Eq.(4.2) must 
sufficiently be canceled among $\lambda_1$, $\lambda_2$ 
and $\lambda_3$, or all of $\lambda_i$ must be 
of the order of $\lambda_K\times 10^{-2}$. 
For a special case in which the Sumino mechanism holds 
exactly, i.e. $\xi=0$, the condition (3.1) is required
in addition to the relation(2.12).
The simultaneous equations (2.12) and (3.1) require the 
numerical ratios among $\lambda_i$ as (3.4).
It does not seems that the numerical results (3.4) are 
understood from a simple model.  

In contrast to the case $B=0$, the case $A=0$ leads 
to an interesting constraint for $\lambda_i$, (3.8),
i.e. $\lambda_2/\lambda_4=2/3$ for a case with 
$\lambda_3=0$, and $\lambda_3/\lambda_4=2/3$ for a case 
with $\lambda_2=0$. 
Such cases are likely in building an explicit model.
[However, we will need an additional ansatz which gives
$A=0$, Eq.(2.6).] 
It seems to be worthwhile searching a model with 
$A=0$ even if we pay a cost of such an additional 
ansatz.

The conditions A and B are dependent on a structure 
of $V_K$.
For example, in the original Sumino model \cite{Sumino09JHEP}, 
the relation (1.1) is exactly derived at an energy scale
at which U(9) flavor symmetry is broken into U(3)$\times$O(3).
The potential $V_S$ effectively appears at a level of 
U(3)$\times$O(3).  
In his model, only the relation C is satisfied, but the
conditions A and B are free.
Even in such a case, the general formula (2.7) is exactly 
required, so that the use of the formula (2.7) will be useful
for building a model of leptons.

\vspace{3mm}

\centerline{\large\bf Acknowledgments}

The author would like to thank Y.~Sumino for valuable and 
helpful conversations. 
This work is supported by the Grant-in-Aid for
Scientific Research (C), JSPS, (No.21540266).


\end{document}